\documentclass[conference,letterpaper]{IEEEtran}

\usepackage[utf8]{inputenc}
\usepackage[T1]{fontenc}
\usepackage{url}              
\usepackage{cite}             

\usepackage[cmex10]{amsmath}  

\usepackage{mathpazo}
\usepackage{times}
\usepackage{amsmath}
\usepackage{amsfonts}
\usepackage{latexsym}
\usepackage{amssymb}
\usepackage{cite}
\usepackage{upref}
\usepackage{theorem}
\usepackage{graphicx}
\usepackage{psfrag}
\usepackage{color}

\newcommand{\err}{\varepsilon_{\tt code}}

\usepackage{mleftright}       
\mleftright                   

\usepackage{graphicx}         
\usepackage{booktabs}         

\newtheorem{proposition}{Proposition}

\newtheorem{claim}{Claim}

\newtheorem{theorem}{Theorem}
\newtheorem{lemma}{Lemma}
\newtheorem{definition}{Definition}
\newtheorem{corollary}[theorem]{Corollary}

\DeclareMathOperator*{\nn}{\nonumber}
\DeclareMathOperator{\E}{\mathbb{E}}
\DeclareMathOperator{\cP}{\mathcal P}
\DeclareMathOperator{\cX}{\mathcal X}
\DeclareMathOperator{\bX}{\bf X}
\DeclareMathOperator{\bx}{\bf x}
\DeclareMathOperator{\by}{\bf y}

\DeclareMathOperator{\cY}{\mathcal Y}
\DeclareMathOperator{\cS}{\mathcal S}
\DeclareMathOperator{\cR}{\mathcal R}
\DeclareMathOperator{\cW}{\mathcal W}
\DeclareMathOperator*{\bp}{\bf p}
\DeclareMathOperator*{\bs}{\bf s}
\DeclareMathOperator*{\bpp}{{(\bf p)}}

\DeclareMathOperator*{\bpps}{{(\bf p, \bf s)}}
\DeclareMathOperator*{\CR}{\mathbf{CR}}
\DeclareMathOperator*{\regret}{\mathbf{Regret}}
\newcommand{\cC}{{\cal C}}

\begin{document}

\title{Competitive Analysis of Arbitrary Varying Channels}

\author{%
  \IEEEauthorblockN{Michael Langberg}
  \IEEEauthorblockA{Department of Electrical Engineering\\
                    University at Buffalo\\
                    Buffalo, NY, USA\\
                    Email: mikel@buffalo.edu}
  \and
  \IEEEauthorblockN{Oron Sabag}
  \IEEEauthorblockA{School of Computer Science and Engineering\\
  The Hebrew University of Jerusalem\\ 
                    Jerusalem, Israel\\
                    Email: oron.sabag@mail.huji.ac.il}
}

\maketitle

\begin{abstract}
Arbitrary varying channels (AVC) are used to model communication settings in which a channel state may vary arbitrarily over time.
Their primary objective is to circumvent statistical assumptions on channel variation. Traditional studies on AVCs optimize rate subject to the worst-case state sequence. While this approach is resilient to channel variations, it may result in low rates for state sequences that are associated with relatively good channels. This paper addresses the analysis of AVCs through the lens of competitive analysis, where solution quality is measured with respect to the optimal solution had the state sequence been known in advance.
Our main result demonstrates that codes constructed by a single input distribution do not achieve optimal competitive performance over AVCs. This stands in contrast to the single-letter capacity formulae for AVCs, and it indicates, in our setting, that even though the encoder cannot predict the subsequent channel states, it benefits from varying its input distribution as time proceeds. 
\end{abstract}

\section{Introduction}
The arbitrarily varying channel (AVC) model, introduced by Blackwell, Breiman, and Thomasian \cite{blackwell_capacities_1960}, captures communication over a collection of memoryless channels, $\cW=\{W_s(y|x)\}_{s \in {\mathcal S}}$, where in each time instance $i$ the channel state that determines the channel in use may be arbitrarily chosen. The AVC model is broad in nature, and in its full generality can capture the setting in which the state sequence may depend on the transmitted codeword or may have a cost/type constraint. 
As such, the AVC model captures both adversarial and random noise models. The majority of previous studies on AVCs address the case in which the state does not depend on the transmitted codeword, e.g., \cite{ahlswede_elimination_1978,CsiszarN:88constraints,CsiszarN:91gavc,CsiszarN:88positivity}.
Traditionally, success criteria for communication in the context of AVCs  require the design of a single coding scheme that allows communication at a fixed rate no matter which state sequence is realized; this approach forces code rates to be matched to the worst-case channel conditions. 
The fixed-rate setting is complemented by variable-rate performance criteria that no longer guarantee the delivery of a fixed rate under all channel state sequences; instead, they allow the rate to vary with the channel states in operation.
{\em Rateless} codes, introduced in \cite{luby2002lt,mackay2005fountain,shokrollahi2006raptor}, achieve variable-rate coding by allowing the effective blocklength to vary with the channel state.

In this work, we consider rateless codes for AVCs. In particular, the message length is fixed in advance and the communication length is a random variable that depends on channel outputs, i.e., the block length is determined by a stopping time based on the channel output filtration. 
Rateless codes for the AVC  model have seen a number of studies over the last decade.
The majority of studies involve feedback\cite{shayevitz2005communicating,eswaran2007using, eswaran2009zero, shayevitz2009achieving,woyach2012comments,lomnitz2011communication,lomnitz2012communication, Blits:12,lomnitz2013universal,joshi2022capacity} and are less relevant to the work at hand.
In the context of rateless codes for AVCs without feedback, prior works include  
\cite{draper2009rateless,sarwate_robust_2008,sarwate2010rateless} that study coding solutions and effective rate in the setting in which the decoder has full or partial access to state information.
Beyond the assumptions on decoder state information (which is also central to our study as well), the major difference between the works above and the work at hand lies in the quality measures of the solutions suggested - the former, for a fixed input distribution, seek decoding rules that minimize the expected decoding time given the state sequence at hand, while the work at hand seeks coding solutions with a {\em competitive} quality guarantee.


Namely, in this work, we study rateless coding technologies for communication over AVCs through the lens of competitive analysis. In competitive analysis, one compares the achievable rates of solutions in which state sequence is not available to the encoder and decoder with those in which state information is known in advance to all parties. The objective is to design communication schemes that achieve rates that are {\em close} to that achievable when the state sequence is known in advance. Common metrics to compare these two rates include the {\em competitive-ratio} that measures the ratio between the (expected) rates achievable in the case of limited state knowledge and that with full state knowledge, and {\em regret} that measures the difference between the former and latter rates described above. The design of communication schemes with a competitive ratio approaching 1 (or with regret approaching 0) guarantees that even in the face of uncertainty, the quality of communication matches the best possible under the given conditions. A competitive ratio $\alpha$ that is bounded away from $1$ acts as a quality measure for the communication scheme at hand, guaranteeing that no matter what state sequence is realized, the achievable rate is guaranteed to be within an $\alpha$ multiplicative ratio of the best possible.

The remainder of the paper is structured as follows.
In Section~\ref{sec:setting}
we present our model and problem definition. 
We then formulate the main question addressed in this work in Section \ref{sec:question}.
Our results are stated in Section \ref{sec:main}.

\section{Model and problem definition}\label{sec:setting}
Let $\cX$, $\cY$, and $\cS$ denote finite alphabets of the channel input, output and state, respectively. Consider a message $M$ uniformly distributed over $\mathcal M \triangleq [1:2^k]$ and communication over channels taken from a family of discrete memoryless channels $\mathcal W = \{W_s(y|x)\}_{s\in\mathcal S}$. The channel at time $i$ is determined by the state $s_i$ as $W_{s_i}(y|x)$. The state sequence does not depend on the message and should be viewed as a deterministic sequence that is chosen arbitrarily.

The message length $k$ is not parameterized with a blocklength or rate. Instead, we consider rateless communication where the fixed-size message should be decoded with the least number of channel uses. That is, the decoder observes the stream of channel outputs and decides at each time if it wants to proceed with communication or to abort it and to decode the message. We proceed to formally define rateless codes.

\vspace{1mm}\noindent
{\bf $\bullet$ Rateless codes over AVCs:}
For a fixed message length $k$, a rateless code $\mathcal C_k\triangleq(E,\{D_i\}_{i=1}^\infty,\{H_i\}_{i=1}^\infty)$ is defined by three {deterministic} mappings. The encoder is defined by the mapping $E: \mathcal M \to \mathcal X^\infty$. The second mapping is a sequence of decoder-decision functions $H_i: \mathcal Y^i \to \{0,1\}$ for  $i \ge1$; if the decoder decodes at time $i$, it sets $H_i(Y^i)=1$ and then the message is decoded with the last mapping $D_i: \mathcal Y^i \to  \mathcal M$. In all time instances prior to the decoding, the decoder simply sets $H_i=0$. 

The stopping time is defined as $\tau_k = \min_i \{i: H_i(Y^i) =1 \}$. For a given code and a state sequence $\bs\in \mathcal S^\infty$, the average probability of error is $ P_e(\bs)\triangleq \Pr(M \neq  D_{\tau_k}(Y^{\tau_k}) \mid \bs)$, and $P_e = \max_{\bs} P_e(\bs)$ denotes the maximal error among all state sequences. The expected decoding time for $\bs$ is~$\tau_k(\bs)= \E[\tau_k\mid\bs]$ (here, expectation is taken over the channel and message). For rateless codes, the effective rate is typically measured as $\frac{k}{\tau_k(\bs)}$, e.g, \cite{burnashev1976data}. 

\vspace{1mm}\noindent
{\bf $\bullet$ Competitive analysis:}
To define our competitive metrics, as our baseline, we consider codes where the encoder and the decoder mappings have access to the state sequence $\bs \in \cS^\infty$. Specifically, it is a rateless code $\mathcal C_k$ as above, but the mappings $(E,\{D_i\},\{H_i\})$ depend on the state sequence. We denote such codes with $\mathcal C^\ast_k$. 

The main idea is to compare the stopping time  $\tau_k(\bs)$ of a code $\cC_k$ with the stopping time $\tau_{k,\epsilon}^*(\bs)$ of an optimal, {\em clairvoyant}, scheme $\cC_k^*$ specifically designed for $\bs$. Setting $\tau_{k,\epsilon}^*(\bs)=\inf\limits_{\mathcal C^*_k:P_e\le\epsilon } \tau_k(\bs)$, the competitive ratio is defined as 
\vspace{-3mm}
\begin{align}\label{eq:def_CR_finite}
\CR(k,\epsilon)&= 
\sup_{\mathcal C_k:P_e\le\epsilon }\inf_{\bs\in\mathcal S^\infty} \frac{\tau^\ast_{k,\epsilon}(\bs)}{\tau_k(\bs)}.
\vspace{-3mm}
\end{align}

The competitive ratio guarantees the largest multiplicative measure of quality with respect to the optimal code. That is, for any $\bs$, whether this sequence implies low or high $\tau^*_{k,\epsilon}(\bs)$, the resulted stopping time of the oblivious code (that is not designed with the knowledge of $\bs$) should be the closest possible to the optimal stopping time.


For some applications, e.g., delay-sensitive systems \cite{hu2018swipt,hu2020throughput,ghanami2020performance,agrawal2022finite,popovski2018wireless,xURLLC,mahmood2023ultra}, one may be interested in additive bounds on the normalized delay or the (effective) rate when compared to an optimal code given state knowledge, defined here as the regret:
\begin{align}\label{eq:def_regret_finite}
\regret(k,\epsilon)&= 
\inf_{\mathcal C_k:P_e\le\epsilon }\sup_{s\in\mathcal S^\infty} \left( \frac{k}{\tau^\ast_{k,\epsilon}(\bs)} - \frac{k}{\tau_k(\bs)}\right).
\end{align}
A family of channels $\mathcal W=\{W_s\}_{s \in \mathcal S}$ is said to be $\alpha(k,\epsilon)$-competitive if $\CR(k,\epsilon) \ge \alpha(k,\epsilon)$.
Similarly for the regret objective, $\mathcal W$ allows a regret of  $\rho(k,\epsilon)$ if $\regret(k,\epsilon) \le \rho(k,\epsilon)$.

In the asymptotic setting, which is the focus of this work, a family of channels $\mathcal W=\{W_s\}_{s \in \mathcal S}$ is said to be $\alpha$-competitive if there exists a sequence $\epsilon_k\to0$ such that 
    $\limsup_{k\to\infty} \CR(k,\epsilon_k) \ge \alpha.$
The supremum over all such $\alpha$ is defined as the optimal competitive ratio (also referred to as the {\em competitive AVC capacity}) and is denoted by $\CR$. Namely,
\begin{align}
\label{eq:CR_inf}
\CR&\triangleq \limsup_{\substack{k\to\infty, \epsilon_k\to0 }} \CR(k,\epsilon_k).
\end{align}
Similarly, a regret of $\rho$ is achievable if there exists $\epsilon_k\to0$ such that $\liminf_{k\to\infty} \regret(k,\epsilon_k) \le \rho$,
and the infimum over such $\rho$ is denoted by~$\regret\triangleq\liminf_{\substack{k\to\infty \\ \epsilon_k\to0 }} \regret(k,\epsilon_k).$
In the remainder of this work, we focus on the $\CR$ quality measure; however,
the results presented here apply also to $\regret$. 

\section{Main question}\label{sec:question}
The main objective towards the design of practical codes that achieve the optimal competitive ratio is a single-letter characterization of $\CR$.
Single-letter characterizations provide efficient means to determine fundamental quality limits and, more importantly, they often provide a simple structure for optimal code design.
In a previous work of the authors, \cite{LangbergSabag24}, a single-letter characterization for the competitive analysis of the {\em compound} channel (in which the state sequence is unknown but does not change over time) was presented.
In this work, in which we address AVCs, we do not derive a single-letter characterization for $\CR$; however, we make significant steps towards understanding this ultimate goal.

In this work, we seek to understand the structure of optimal codes for competitive metrics.
Specifically, we ask if, similar to classical results in the context of communication, coding schemes designed and governed by a {\em single} optimizing distribution are optimal in the competitive setting as well.
We note that, in the traditional study of AVCs, the capacity (up to symmetrizability) is characterized as $\max_{p(x)}\min_{p(s)} I(X;Y)$, e.g., \cite{LapidothN:98survey}, by a single input distribution. Here $(X,Y)$ are distributed according to $\sum_s p(x)p(s)W_s(y|x)$.
The same holds for the rateless study of such AVCs when one wishes to minimize the worst-case decoding time, 
with \cite{draper2009rateless} and without \cite{kosut2018finite} decoder state information.

For the compound setting, \cite{LangbergSabag24} show that single-distribution codes are {\em suboptimal} in the context of competitive analysis.
Namely, for competitive quality measures in the compound setting, the encoder, even without additional feedback knowledge, may be required to modify encoding statistics as time goes by. 
This follows, roughly stated, since a competitive encoder at any given time must optimize performance over a monotonely shrinking set of channel states. 
Initially, the encoder must act with all potential states in mind.
However, as time goes by, certain channel statistics, if realized, would have already allowed successful decoding, and thus no longer need to be considered in the encoder's optimization.
In this work, we ask whether this sub-optimality applies to the competitive analysis of the {\em time-varying} AVC model as well.
In AVCs, the encoder can no longer rule out future channel states based on prior assumptions. 
Why then should an AVC encoder change its encoding statistics as time goes by?
Rather surprisingly, we show that, even without additional feedback knowledge, the optimal AVC encoder may be required to modify its statistics as time goes by.
In what follows, we introduce additional notation that allows us to formally state and then prove our main result.


We address the number of input distributions needed in code design to achieve an optimal competitive ratio. Formally, a uniform message $M$ and an encoder mapping $E$ induce a distribution on the infinite ensemble of channel inputs $\cP(\cX^\infty)$. 
Computing the marginal distribution at each time step, we obtain a product distribution $p(x_1)p(x_2)\cdots \in \cP(\cX)^\infty$.
The product distribution does not characterize the code at hand (as different inputs may depend on each other) - but will suffice for our purposes.
We ask how many different distributions in $\cP(\cX)$ are needed to achieve an optimal competitive ratio. 

We define product distributions that alternate at $\ell-1$ points as follows:
\begin{definition} \label{def:prob}
    The set $\cP_\ell \subset \cP(\cX)^\infty$ includes all product distributions $\bp \in \cP(\cX)^\infty$ with $\bp=p_1^{n_1},p_2^{n_2},\dots,p_{\ell-1}^{n_{\ell-1}},p_{\ell}^\infty$ where $\{p_1,\dots,p_\ell\} \subset \cP(\cX)$ and $n_1,\dots,n_{\ell-1}$ are non-negative integers.
\end{definition}
For example, if $\ell=1$, the channel input has the same marginal distribution on $X_i$ for all $i\ge1$. 
Complementing Definition~\ref{def:prob}, the competitive ratio $\CR_\ell$ is defined similarly to $\CR$ but when the codes in use are restricted to have product distributions {\em close} to  $\cP_\ell$ (slackness is added to allow the slight variability implied by random code design).
Specifically, for $\delta>0$ and a given $k$, a codebook $E:[2^k]\rightarrow \cX^\infty$ 
is said to be $\delta$-close to $\bp =p_1,p_2,\dots$
if the marginal distributions $\{q_i\}_{i=1}^{\infty}$ corresponding to the $i$'th codeword entry satisfy for every integer $n$ that 
$\sum_{i=1}^n{\tt TV}(p_i,q_i)\leq \delta n$.
Here, ${\tt TV}$ is the total variation distance between distributions.
A codebook $E:[2^k]\rightarrow \cX^\infty$ is said to be $\delta$-close to $\cP_\ell$ if it is $\delta$-close to some $\bp \in \cP_\ell$.
Let $\CR_\ell$ be defined similarly to $\CR$ when restricted to codes with product distributions $\delta$-close to  $\cP_\ell$ for $\delta \rightarrow 0$.
We give a formal definition of $\CR_\ell$ in 
Appendix~\ref{sec:cr_ell}.
We now have that  
\begin{align}
    {\CR}_1 \le {\CR}_2 \le \cdots\le {\CR}.
\end{align}

As stated earlier, traditional results on communication over memoryless channels and AVCs show that there exist optimal encoders (with corresponding product distributions) that lie in $\mathcal P_1$.
For the compound channel, \cite{LangbergSabag24} show, on the one hand, that codes in $\mathcal P_1$ are sub-optimal for competitive measures, but, on the other, that codes in 
 $\cP_{|\mathcal S|}$ are sufficient for competitive optimality.
 The latter, for finite $|\mathcal S|$, enables a single-letter characterization of $\CR$ in the compound setting (as an optimization over $|\mathcal S|$ input distributions).



The work at hand 
asks whether $\cP_1$ suffices to achieve optimal competitive measures on AVCs. We show through an example that {\bf single-distribution codes do not suffice to achieve the competitive capacity $\CR$ for AVCs.} That is, $\CR> \CR_1$.
Even given the worst-case flavor of the AVC model, namely, given the fact that at every time step any channel can come next, the fact that for certain channel sequences, decoding can be done before others, leads to the realization that the encoder may benefit from different behaviors as time passes.

\section{Main Result}\label{sec:main}
In this section, we present our main result and the main steps to establish it. \begin{theorem}\label{th:main}
Single input distributions do not achieve the optimal competitive ratio in AVCs. That is, $\CR_1 < \CR$.
\end{theorem}
The assertion implies that there exists AVCs for which any single-letter characterization of the optimal competitive ratio should include at least two channel input distributions. 
Whether there is an upper bound to the number of input distributions sufficient for optimal competitive codes, i.e., whether there exists $\ell <\infty$  such that ${\CR}_\ell = \CR$, is left open in this work.

To prove Theorem \ref{th:main}, we show for a particular channel family the following chain of inequalities. 
\begin{align}\label{eq:main_numbers}
 1/3&\stackrel{(a)}= {\CR}_1 < 11/24\stackrel{(b)}\le {\CR}_2 \le \CR\stackrel{(c)}\le 1/2.
\end{align}
To prove Theorem \ref{th:main}, it is sufficient to prove steps $(a)-(b)$ in \eqref{eq:main_numbers}. These steps are proved in Section \ref{sec:proof_main}. Furthermore, we prove step $(c)$ since this bound is quite close to the lower bound of step $(b)$, and its proof method may be of independent interest.

The channel family we study to establish \eqref{eq:main_numbers} is depicted in Fig. \ref{fig:avc_example}, and is discussed in detail in Section \ref{sec:proof_main}. 
In the particular example we study, the channel state can be determined by the channel output. 
Thus, in our achievability schemes for Theorem~\ref{th:main}, we consider rateless codes for the AVC setting in which the decoder has full state information (DSI).
Consequently, we employ Theorem~\ref{th:multi} stated below in the proof of Theorem~\ref{th:main}.
While the proof of Theorem~\ref{th:multi} may follow from various modifications in prior works that study rateless AVC achievability in the presence of DSI \cite{draper2009rateless,sarwate_robust_2008,sarwate2010rateless}, as our model is slightly different, a self-contained proof of Theorem~\ref{th:multi} appears in 
Appendix~\ref{app:ach_proof}.
In what follows, we consider codes $\mathcal C^{\tt DSI}_k=(E,\{D_i\}_{i=1}^\infty,\{H_i\}_{i=1}^\infty)$ with decoder state information (DSI). That is, the decoder mappings depend causally on the channel outputs and states.

\begin{theorem}\label{th:multi}
Let $\delta>0$.
Let $k$ be sufficiently large.
Let $\ell<\infty$.
Let $\bp = p_1^{n_1},p_2^{n_2},\dots,p_{\ell-1}^{n_{\ell-1}},p_{\ell}^\infty \in \cP_\ell$.
Let $\cW=\{W_s\}$ be a channel family for which 
$\min_{i,s}I_s(X,Y)>0$, where for state $s$ and index $i$, $(X_i,Y)$ is distributed according to $p_i(x)W_s(y|x)$. 
For $\bs \in \cS^\infty$ let
$$
\tau^{({\bf p})}_{k}({\bf s}) = \arg\min_\tau{\{\sum_{t<\tau}I_{s_t}(
X_{t},Y) \geq k\}}.
$$
Then, there exists a code $\mathcal C^{\tt DSI}_k=(E,\{D_i\}_{i=1}^\infty,\{H_i\}_{i=1}^\infty)$ that is $\delta$-close to $\bp$ with decoding error probability $\epsilon_k \leq \frac{1}{k}$ and competitive ratio at least
$$
(1-\delta) \inf_{{\bf s} \in\mathcal S^\infty} \left(\frac{\tau^*_k({\bf s})}{\tau^{({\bf p})}_k({\bf s})}\right),
$$
where 
$\tau_k^*(\bs) = \arg\min_\tau\{\sum_{t<\tau}C(s_t)\ge k\}$ with $C(s) = \max_{p(x)}I_s(X;Y)$.

\end{theorem}



The intuition behind Theorem \ref{th:multi} is that a decoder with state information can successfully decode the message once the accumulated mutual information between the channel input and output surpasses the message entropy. 
The optimal stopping time, $\tau_k^*(\bs)$, follows in a similar way when the encoder can use the distribution achieving capacity for each state $s_t$.

\section{Proof of Theorem~\ref{th:main}}\label{sec:proof_main}
In this section, we present a family of channels such that the best competitive ratio with a single input distribution is $\CR_1=1/3$, where one can obtain $\CR_2 \geq 11/24\approx0.46$ if two input distributions are allowed. This establishes the proof of Theorem \ref{th:main}. The proof of step $(c)$ in \eqref{eq:main_numbers}, $\CR\le 1/2$, appears in 
the appendix.

\begin{figure}[t]
	\vspace{-1mm}
 \includegraphics[scale=0.6]{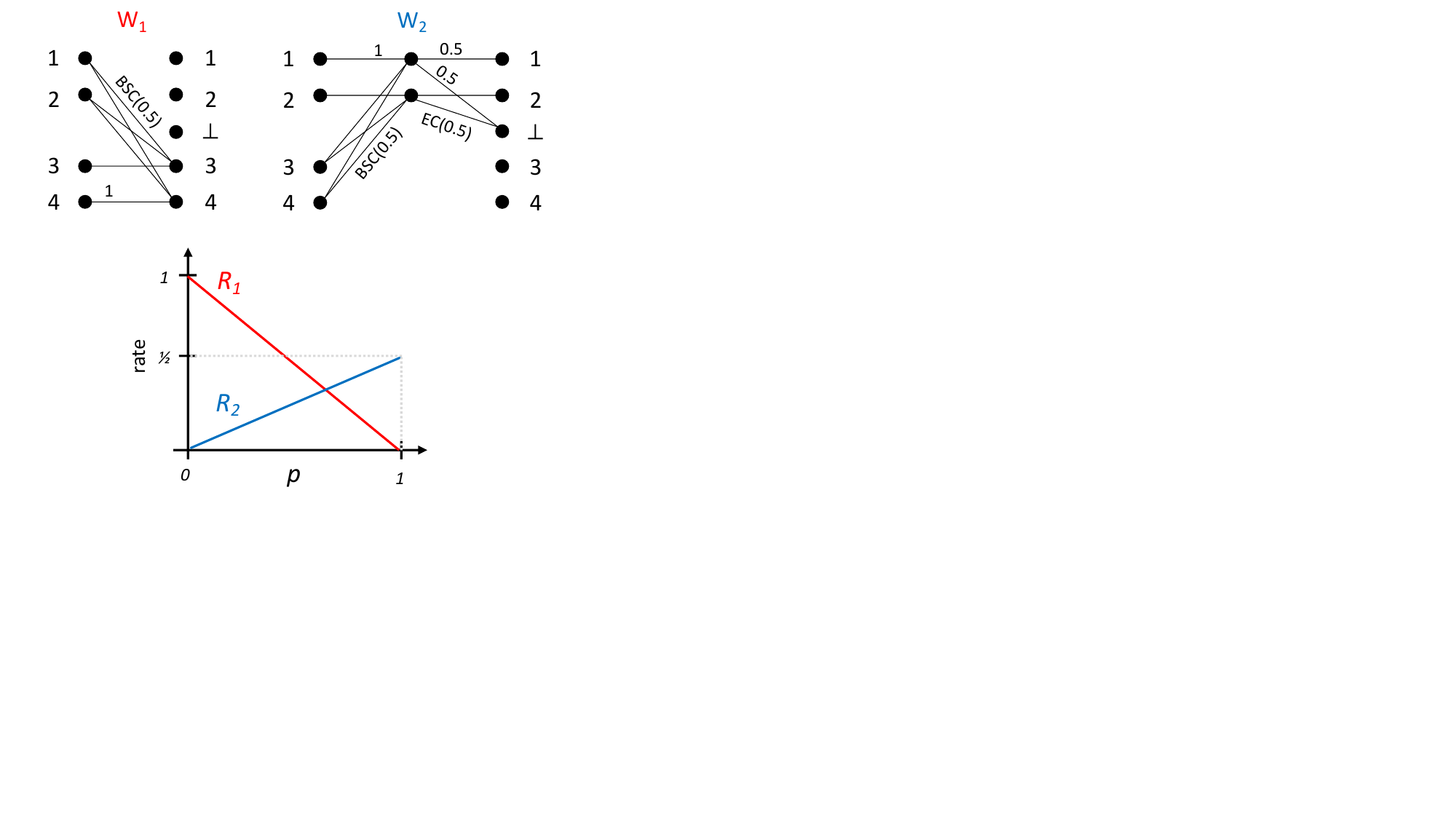}
	\vspace{-49mm}
	\caption{The channels $W_1$ and $W_2$ (top). $R_1(p)$ vs. $R_2(p)$ (bottom). }
 	\label{fig:avc_example}
\end{figure}

The channel family consists of two channels and is depicted in Figure~\ref{fig:avc_example}. Let $\mathcal X= \{1,2,3,4\}$ and $\mathcal Y=\{1,2,3,4,\perp\}$. In the first channel, $W_1$, if $x \in \{3,4\}$ the output is $y=x$, and if $x \in \{1,2\}$ the output $y$ has a uniform distribution on $\{3,4\}$.

In the second channel, $W_2$, if $x \in \{1,2\}$ the output $y$ is $y=x$ with probability $0.5$ and $y=\perp$ with probability $0.5$. For $x \in \{3,4\}$, channel $W_2$ sets $y$ to be uniform in $\{1,2\}$ with probability $0.5$ and $y=\perp$ with probability $0.5$. The channel $W_2$ is represented as the concatenation of two channels, a first channel followed by an erasure channel (EC). We will use the fact that the channel capacities are $C_1=1$ and $C_2=0.5$, respectively.

The example is a modified version of the ``bilingual speaker'' example from \cite{shulman2000static,Shulman:09}, and was also studied in our prior work on the competitive analysis of compound channels \cite{LangbergSabag24}. Note that the encoder is uncertain about the state sequence, but the decoder can deduce the channel state from the channel output since $y\in\{3,4\}$ if and only if the first channel is used. Thus, we can utilize Theorem \ref{th:multi} that addresses code analysis in the setting of DSI.

For code design, by symmetry, it is sufficient to consider for $p \in [0,1]$ input distributions $X$ that satisfy 
\begin{align}\label{def:input}
  \Pr(X=1)&=\Pr(X=2)=\frac{p}{2}\nn\\
  \Pr(X= 3)&= \Pr(X=4)=\frac{1-p}{2}.  
\end{align}
The achievable rates for $W_1$ and $W_2$ for single-distribution codes governed by such $X$ equal $R_1(p)=1-p$ and $R_2(p)=\frac{p}{2}$ as depicted in Figure~\ref{fig:avc_example}.

By Theorem~\ref{th:multi}, there exists a communication scheme that allows successful decoding once the cumulative mutual information exceeds the message length. Roughly speaking, decoding is successful after 
\begin{align}\label{eq:proof_accumulate}
\tau^{({\bf p})}_k({\bf s}) =  \arg\min_\tau{\{\sum_{t<\tau}I_{s_t}(
X_{t},Y)\geq k\}},    
\end{align}
where $X_t$ corresponds to the distribution of the input at time $t$ and can be parameterized by $p\in[0,1]$. We will also utilize the optimal decoding time $\tau_k^*(\bs)$ defined in Theorem \ref{th:multi}.

\subsection{Proof of $\CR_1 = 1/3$:} 
A lower bound is achieved by setting the channel input distribution in \eqref{def:input} with $p=2/3$. This implies that the mutual information under both channels is $I_1(X;Y) = I_2(X;Y) = 1/3$ and therefore $\tau_k^{({\bf p})}(\bs) = 3k$ for all $\bs$. The competitive ratio is thus bounded, for any $\delta>0$, by $\CR\ge (1-\delta)\lim_{k\to\infty}\inf_{\bs\in \mathcal S^\infty} \frac{\tau_k^*(\bs)}{3k} \geq \frac{1-\delta}{3}$ where the equality follows by setting the prefix of $\bs$ to $1^k$ since it minimizes the optimal decoding time.

To prove the upper bound on ${\CR}_1$, it suffices to consider the inequality 
\begin{align}
\label{eq:lower}
        & \lim_{k\to\infty}\max_{\bp \in \mathcal P_1} \min_{\bs\in \mathcal S^{\infty}} \left(\frac{\tau^*_k({\bf s})}{\tau^{({\bf p})}_k({\bf s})}\right)\le \lim_{k\to\infty}\max_{\bp \in \mathcal P_1} \min_{\bs\in  \widehat{\mathcal S}^{\infty}} \left(\frac{\tau^*_k({\bf s})}{\tau^{({\bf p})}_k({\bf s})}\right), 
\end{align}
for $\widehat{\mathcal S}^\infty \subseteq \mathcal S^\infty$ of our choice. 
Bounding $\CR_1$ by the left expression in \eqref{eq:lower} roughly follows from standard capacity bounds, e.g., \cite[Section VI]{LangbergSabag24}.
For the right expression, 
we analyze the setting $\widehat{\mathcal S}^\infty = \{2^\infty, 1^k2^\infty\}$. 

For the first sequence $\bs_1=2^\infty$, the optimal decoding time is $\tau_k^*(\bs_1) = 2k$. For this sequence, utilizing a code with a single distribution $p\neq 0$, we have $\tau^{(\bp)}_k(\bs_1) = 2k/p$ (We can assume $p\neq0$ since otherwise the decoding time for $\bs_1$ is infinite). For the second sequence, $\bs_2=1^k2^\infty$, we have $\tau_k^*(\bs_2) = k$ and, for $p\neq0$,  $\tau^{(\bp)}_k(\bs_2) = 3k$. Combining the two ratios that correspond to the different sequences and taking limits over $k$ we obtain 
\begin{align}
    {\CR}_{1}&\le \max_{p\in(0,1]} \min \{p,1/3\} = 1/3.
\end{align}

\subsection{Proof of $\CR_2\ge 11/24$:}
The prove the assertion, we present a family of codes with two input distributions and analyze their performance. We consider codes that utilize $p_1\le 2/3$ until time $t k$ and $p_2=2/3$ afterwards. Without loss of optimality, we consider $1\le t\le2$ since the minimal and maximal optimal stopping times are within this range. The optimal parameters will be shown to be $p_1=10/33$ and $t=3/2$.


For any state sequence $\bs \in \mathcal S^\infty$, let its optimal decoding time be $\tau_k^*(\bs)=rk$ where $r \in [1,2]$ and depends on the underlying state sequence.\footnote{With some abuse of notation, throughout the analysis, we omit sub-linear terms of $k$ that do not affect the competitive ratio analysis.} Let $\alpha$ be the fraction of $1's$ in $\bs^{rk}$, and $(1-\alpha)$ be the fraction of 2's in $\bs^{rk}$. By the accumulation of mutual information, it holds that $k = \alpha r k C_1 + (1-\alpha) r k C_2$ where $C_1=1, C_2=0.5$ are the capacities of the channels, respectively; or in other words $\alpha r=2-r$. Let $\beta\ge1$ be a constant that is independent of $\bs$ and satisfies $\tau_k(\bs) \ge \beta rk$ for all $\bs$. Our objective is to find the least $\beta$ that satisfies the  inequality for all $\bs$ since it directly implies $\CR_2\ge \frac1{\beta}$.

We proceed to the analysis broken to sub-cases based on the value of $r\in[1,2]$.



{\bf Case 1: $r \in [1,t]$.} The range of $r$ implies the ordering $1\le r\le t \le \beta r$.\footnote{The case $\beta r\le t$ is not possible; this can only occur if $t\ge\beta\ge1$, but this contradicts our assumption that $t\le2$ combined with our upper bound $\CR\le 1/2$.} That is, the alternating distribution point $tk$ lies between the stopping time and the optimal stopping time. 

Let $\bs_{rk}$ be the prefix of $\bs$ up to length $rk$. By our assumption that $p_1\le2/3$, we can consider $\bs_{rk}2^\infty$ instead of $\bs$ since this sequence has the largest stopping time. We have by the accumulation of the mutual information in \eqref{eq:proof_accumulate}
\begin{align}
    k&\ge (1-\alpha)rk\frac{p_1}{2} + \alpha rk (1-p_1) + (tk-rk)\frac{p_1}{2} + \frac{\tau_k(\bs) - tk}{3}\nn\\
    &\ge (1-\alpha)rk\frac{p_1}{2} + \alpha rk (1-p_1) + (tk-rk)\frac{p_1}{2} + \frac{\beta r k- tk}{3},\nn
\end{align}
which implies $\beta\le \frac{18p_1 + 6r + 2t - 9p_1r - 3p_1t - 6}{2r}$. This inequality holds for all state sequences such that their corresponding $r$ satisfies $r \in [1,t]$, so we proceed to maximize the upper bound by maximizing it over $r$.

The bound is increasing (in $r$) if $-18p_1 + 2t - 3p_1t - 6\ge0$ and is decreasing otherwise. That means that if the condition is satisfied, the bound is maximizes at $r^* = t$ resulting in $\beta\le \frac{9p_1 + 4t - 6p_1t - 3}{t}$. If the condition is not satisfied, the upper bound is maximized at $r^*=1$ resulting in $\beta\le \frac{9p_1  + 2t  - 3p_1t }{2}$. It is easy to verify that the condition is not satisfied for $t\in[1,2]$, and so we will use the second bound.

{\bf Case 2: $r \in [t,\frac{t+2}{2}]$.} The range of $r$ implies that the number of 2's can be contained within $[0,tk]$. That is, $r\le \frac{t+2}{2} \iff (1-\alpha)r k \le tk$. This now implies that the worst sequence (for this case) is the one starting with $2$'s, which in turn implies that: 
\begin{align}
    1&\ge (1-\alpha)r\frac{p_1}{2} + (t - (1-\alpha)r) (1-p_1)  + \frac{\beta r - t}{3},
\end{align}
which implies $\beta\le \frac{9p_1 + 6r - 2t - 9p_1r + 3p_1t - 3}{r}$.

The bound is increasing if $t + (t+3)(1-3p_1)\ge0$ implying that $r^* = t$ and $\beta\le \frac{9p_1 + 4t - 6p_1t- 3}{t}$, and if the condition is not satisfied, $r^* = \frac{t+2}{2}$ is the maximizer and we have $\beta\le \frac{2t - 3p_1t + 6}{t + 2}$.

{\bf Case 3: $r \in [\frac{t+2}{2},2]$.} In this range, we have $(1-\alpha)r  \ge t$ so that, in the worst state-sequence for this case, the number of 2's occupies $[0,t]$:
\begin{align}
    1&\ge t\frac{p_1}{2} + \frac{\beta r - t}{3}.
\end{align}
This implies $\beta\le \frac{2t - 3p_1t + 6}{2r}$ and is minimized at $r^* = \frac{t+2}{2}$ so that $\beta\le \frac{2t - 3p_1t + 6}{t+2}$.

\textbf{Summary:} We need to combine the different cases and choose the largest upper bound among these. If the condition of Case $2$ is satisfied, we obtain
{\small{
\begin{align}
    \beta \le \min_{t\in[1,2],p_1\le2/3}\max & \left\{\frac{9p_1  + 2t  - 3p_1t }{2}, \frac{9p_1 + 4t - 6p_1t- 3}{t} ,\right.\nn\\
    & \ \ \ \ \ \ \left.\frac{2t - 3p_1t + 6}{t+2}\right\}.
\end{align}
}}
The second term is dominated by the maximum between the first and third terms for all $p,t$ in the range. 
By comparing the first and third terms we obtain $p_1^*=\frac{2t^2 - 12}{3t^2 - 9t - 18}$ giving 
\begin{align}
    \beta &\le \min_{t\in[1,2]} \frac{18}{- t^2 + 3t + 6}.\nn
\end{align}
The upper bound is minimized at $t=3/2$, implying in turn that $p_1^*=10/33$. To conclude, the optimized bound is $\beta\le \frac{24}{11}$ and proves $\CR_2\ge \frac{11}{24}\approx 0.458$. Note that if the condition of Case $2$ is not satisfied, we necessarily have a greater bound since we have the same optimization as before but the minimum over $t$ is constrained.

\section{Conclusions}
In this work, we study the competitive analysis of AVCs in the rateless setting. Unlike traditional solutions for AVCs, we find that codes using a single input distribution fall short of achieving optimal competitive performance. This emphasizes the necessity of encoding technologies that adapt the input distribution over time, even in the absence of feedback or any knowledge about subsequent channel states. A single-letter expression for the optimal competitive ratio is left open in this work. In particular, an upper bound to the number of input distributions sufficient for optimal competitive codes is subject to future studies.



\IEEEtriggeratref{20}
\bibliographystyle{unsrt}
\bibliography{Bib/online_rateless, Bib/prop}

\begin{thebibliography}{10}

\bibitem{blackwell_capacities_1960}
D.~Blackwell, L.~Breiman, and A.~J. Thomasian.
\newblock The capacities of certain channel classes under random coding.
\newblock {\em The Annals of Mathematical Statistics}, pages 558--567, 1960.

\bibitem{ahlswede_elimination_1978}
R.~Ahlswede.
\newblock Elimination of correlation in random codes for arbitrarily varying
  channels.
\newblock {\em Z. Wahrsch. Verw. Gebiete}, 33:159--175, 1978.

\bibitem{CsiszarN:88constraints}
I.~Csisz\'{a}r and P.~Narayan.
\newblock Arbitrarily varying channels with constrained inputs and states.
\newblock {\em IEEE Transactions on Information Theory}, 34(1):27--34, 1988.

\bibitem{CsiszarN:91gavc}
I.~Csisz\'{a}r and P.~Narayan.
\newblock Capacity of the {Gaussian} arbitrarily varying channel.
\newblock {\em IEEE Transactions on Information Theory}, 37(1):18--26, January
  1991.

\bibitem{CsiszarN:88positivity}
I.~Csisz\'{a}r and P.~Narayan.
\newblock The capacity of the arbitrarily varying channel revisited :
  Positivity, constraints.
\newblock {\em IEEE Transactions on Information Theory}, 34(2):181--193, 1988.

\bibitem{luby2002lt}
M.~Luby.
\newblock {LT codes}.
\newblock In {\em Proceedings of the 43rd Annual IEEE Symposium on Foundations
  of Computer Science}, pages 271--280, 2002.

\bibitem{mackay2005fountain}
D.~J.~C. MacKay.
\newblock Fountain codes.
\newblock {\em IEE Proceedings-Communications}, 152(6):1062--1068, 2005.

\bibitem{shokrollahi2006raptor}
A.~Shokrollahi.
\newblock Raptor codes.
\newblock {\em IEEE Transactions on Information Theory}, 52(6):2551--2567,
  2006.

\bibitem{shayevitz2005communicating}
O.~Shayevitz and M.~Feder.
\newblock Communicating using feedback over a binary channel with arbitrary
  noise sequence.
\newblock In {\em IEEE International Symposium on Information Theory (ISIT)},
  pages 1516--1520, 2005.

\bibitem{eswaran2007using}
K.~Eswaran, A.~D. Sarwate, A.~Sahai, and M.~Gastpar.
\newblock Using zero-rate feedback on binary additive channels with individual
  noise sequences.
\newblock In {\em IEEE International Symposium on Information Theory (ISIT)},
  pages 1431--1435, 2007.

\bibitem{eswaran2009zero}
K.~Eswaran, A.~D Sarwate, A.~Sahai, and M.~C. Gastpar.
\newblock Zero-rate feedback can achieve the empirical capacity.
\newblock {\em IEEE Transactions on Information Theory}, 56(1):25--39, 2009.

\bibitem{shayevitz2009achieving}
O.~Shayevitz and M.~Feder.
\newblock Achieving the empirical capacity using feedback: Memoryless additive
  models.
\newblock {\em IEEE Transactions on Information Theory}, 55(3):1269--1295,
  2009.

\bibitem{woyach2012comments}
K.~Woyach, K.~Harrison, G.~Ranade, and A.~Sahai.
\newblock Comments on unknown channels.
\newblock In {\em Information Theory Workshop}, pages 172--176. IEEE, 2012.

\bibitem{lomnitz2011communication}
Y.~Lomnitz and M.~Feder.
\newblock Communication over individual channels.
\newblock {\em IEEE Transactions on Information Theory}, 57(11):7333--7358,
  2011.

\bibitem{lomnitz2012communication}
Y.~Lomnitz and M.~Feder.
\newblock Communication over individual channels--a general framework.
\newblock {\em arXiv preprint, arXiv:1203.1406}, 2012.

\bibitem{Blits:12}
N.~Blits.
\newblock Rateless codes for finite message set.
\newblock {\em M.Sc. dissertation, Tel-Aviv University}, 2012.

\bibitem{lomnitz2013universal}
Y.~Lomnitz and M.~Feder.
\newblock Universal communication over arbitrarily varying channels.
\newblock {\em IEEE Transactions on Information Theory}, 59(6):3720--3752,
  2013.

\bibitem{joshi2022capacity}
P.~Joshi, A.~Purkayastha, Y.~Zhang, A.~J. Budkuley, and S.~Jaggi.
\newblock {On the Capacity of Additive AVCs with Feedback}.
\newblock In {\em IEEE International Symposium on Information Theory (ISIT)},
  pages 504--509, 2022.

\bibitem{draper2009rateless}
S.~C. Draper, F.~R. Kschischang, and F.~Brendan.
\newblock Rateless coding for arbitrary channel mixtures with decoder channel
  state information.
\newblock {\em IEEE Transactions on Information Theory}, 55(9):4119--4133,
  2009.

\bibitem{sarwate_robust_2008}
A.~D. Sarwate.
\newblock {\em Robust and adaptive communication under uncertain interference}.
\newblock PhD thesis, University of California, Berkeley, 2008.

\bibitem{sarwate2010rateless}
A.~D. Sarwate and M.~Gastpar.
\newblock {Rateless codes for AVC models}.
\newblock {\em IEEE Transactions on Information Theory}, 56(7):3105--3114,
  2010.

\bibitem{burnashev1976data}
M.~V. Burnashev.
\newblock {Data transmission over a discrete channel with feedback. Random
  transmission time}.
\newblock {\em Probl. Inf. Transm.}, 12(4):250–--265, 1976.

\bibitem{hu2018swipt}
Y.~Hu, Y.~Zhu, M.~C. Gursoy, and A.~Schmeink.
\newblock {SWIPT-enabled relaying in IoT networks operating with finite
  blocklength codes}.
\newblock {\em IEEE Journal on Selected Areas in Communications}, 37(1):74--88,
  2018.

\bibitem{hu2020throughput}
Y.~Hu, Y.~Li, M.~Gursoy, S.~Velipasalar, and A.~Schmeink.
\newblock {Throughput analysis of low-latency IoT systems with QoS constraints
  and finite blocklength codes}.
\newblock {\em IEEE Transactions on Vehicular Technology}, 69(3):3093--3104,
  2020.

\bibitem{ghanami2020performance}
F.~Ghanami, G.~A. Hodtani, B.~Vucetic, and M.~Shirvanimoghaddam.
\newblock {Performance analysis and optimization of NOMA with HARQ for short
  packet communications in massive IoT}.
\newblock {\em IEEE Internet of Things Journal}, 8(6):4736--4748, 2020.

\bibitem{agrawal2022finite}
N.~Agrawal, A.~Bansal, K.~Singh, C.P. Li, and S.~Mumtaz.
\newblock {Finite block length analysis of RIS-assisted UAV-based multiuser IoT
  communication system with non-linear EH}.
\newblock {\em IEEE Transactions on Communications}, 70(5):3542--3557, 2022.

\bibitem{popovski2018wireless}
P.~Popovski, J.~J. Nielsen, C.~Stefanovic, E.~De~Carvalho, E.~Strom, K.~F.
  Trillingsgaard, A.~S. Bana, D.~M. Kim, R.~Kotaba, J.~Park, and R.~B.
  Sorensen.
\newblock {Wireless access for ultra-reliable low-latency communication:
  Principles and building blocks}.
\newblock {\em IEEE Network}, 32(2):16--23, 2018.

\bibitem{xURLLC}
C.~She, C.~Pan, T.~Q. Duong, T.~Q.~S. Quek, R.~Schober, M.~Simsek, and P.~Zhu.
\newblock {Guest Editorial xURLLC in 6G: Next Generation Ultra-Reliable and
  Low-Latency Communications}.
\newblock {\em IEEE Journal on Selected Areas in Communications},
  41(7):1963--1968, 2023.

\bibitem{mahmood2023ultra}
N.~H. Mahmood, I.~Atzeni, E.~A. Jorswieck, and O.~L.~A. L{\'o}pez.
\newblock {Ultra-Reliable Low-Latency Communications: Foundations, Enablers,
  System Design, and Evolution Towards 6G}.
\newblock {\em Foundations and Trends{\textregistered} in Communications and
  Information Theory}, 20(5-6):512--747, 2023.

\bibitem{LangbergSabag24}
M.~Langberg and O.~Sabag.
\newblock Competitive channel-capacity.
\newblock {\em IEEE Transactions on Information Theory}, 2024.

\bibitem{LapidothN:98survey}
A.~Lapidoth and P.~Narayan.
\newblock Reliable communication under channel uncertainty.
\newblock {\em IEEE Transactions on Information Theory}, 44(10):2148--2177,
  1998.

\bibitem{kosut2018finite}
O.~Kosut and J.~Kliewer.
\newblock Finite blocklength and dispersion bounds for the arbitrarily-varying
  channel.
\newblock In {\em IEEE International Symposium on Information Theory (ISIT)},
  pages 2007--2011, 2018.

\bibitem{shulman2000static}
N.~Shulman and M.~Feder.
\newblock Static broadcasting.
\newblock In {\em IEEE International Symposium on Information Theory}, page~23,
  2000.

\bibitem{Shulman:09}
N.~Shulman.
\newblock Communication over an unknown channel via common broadcasting.
\newblock {\em Ph.D. dissertation, Tel Aviv University}, 2003.

\bibitem{Shannon:48communication}
C.~Shannon.
\newblock A mathematical theory of communication.
\newblock {\em Bell System Technical Journal}, 27(3):379--423, 623--656, July
  1948.

\bibitem{CK97}
I.~Csisz\'{a}r and J.~Korner.
\newblock {\em Information Theory: Coding Theorems for Discrete Memoryless
  Systems, 2nd edition}.
\newblock Akademiai Kiado, New York, NY, 1997.

\bibitem{Stam:75}
S.~Z. Stambler.
\newblock {Shannon theorems for a full class of channels with state known at
  the output}.
\newblock {\em Problems of Information Transmission}, 11(4):3--12, (In
  Russian). 1975.

\bibitem{yeung2008information}
R.~W. Yeung.
\newblock {\em Information theory and network coding}.
\newblock Springer Science \& Business Media, 2008.

\end{thebibliography}


\clearpage
\appendices
\section{Definition of $\CR_\ell$}
\label{sec:cr_ell}

Let $\delta>0$.
Let $\ell$ be a positive integer.
Let ${\mathcal{C}}^{(\delta)}_{k,\ell}$ be the set of all rateless codes with message size $k$ that correspond to product distributions ${\bf p}$ that are $\delta$ close to ${\mathcal{P}}_\ell$.
Let,
\begin{align}\label{eq:def_CR_finite_ell}
{\CR}^{(\delta)}(k,\epsilon)&= 
\sup_{\mathcal C_k \in \cC^{(\delta)}_{k,\ell}:P_e\le\epsilon }\inf_{\bs\in\mathcal S^\infty} \frac{\tau^\ast_{k,\epsilon}(\bs)}{\tau_k(\bs)}.
\vspace{-3mm}
\end{align}

Let,
\begin{align}
\label{eq:CR_inf_ell}
{\CR}^{(\delta)}_\ell&=\limsup_{\substack{k\to\infty, \epsilon_k\to0 }} {\CR}^{(\delta)}(k,\epsilon_k).
\end{align}
Finally, let
\begin{align}
\label{eq:CR_inf_ell2}
{\CR}_\ell&=\liminf_{\substack{\delta\to0 }} {\CR}^{(\delta)}_\ell.
\end{align}

\section{Achievability (Proof of 
Theorem~\ref{th:multi})}\label{app:ach_proof}
Let $\delta>0$.
Let $k$ be sufficiently large.
Consider any ${\bf p}=p_1^{n_1},p_2^{n_2}, \dots, p_{\ell-1}^{n_{\ell-1}},p_{\ell}^\infty \in \cP^\infty$ in $\cP_{\ell}$. 
Here, $n_1,\dots,n_{\ell-1}$ may depend on $k$.
In what follows, we show that a rateless code designed at random, i.e., in which the $i$'th codewords entries are drawn independently from the $i$'th entry in $\bp$, has (with high probability) competitive ratio at least
$$
(1-\delta)\inf_{\bs \in {\mathcal S^\infty}}\left(\frac{\tau^*_k({\bf s})}{\tau^{({\bf p})}_k({\bf s})}\right).
$$
\subsection{Overview}
Before diving into the technical proof, we give a rough overview.
In general, in the setting of {\tt DSI} for finite $\cS$, traditional analysis of random code design for memoryless channels \cite{Shannon:48communication} shows for any given fixed ${\bf s}$ that with overwhelming probability $1-\err$ over code construction, the resulting code guarantees a vanishing (in $k$) decoding error once the decoder waits for an appropriate amount of time.
The main question at hand is the decay of $\err$ as $k$ grows.
For several applications, a polynomial or exponential decay of $\err$ in $k$ is sufficient. However, in our case the code should have a vanishing decoding error, no matter which ${\bs}$ is in use. As the number of possible states is exponential, we will require $\err$ to have double-exponential dependence on $k$. 
Analysis using double-exponential (or super-exponential) concentration on code design is rather common in the AVC literature. Once established in our setting, 
combining a few additional ideas with a union bound over all possible ${\bs}$ suffices to conclude that a single code achieves the expression for all states sequence. We proceed to establish the stated concentration on $\err$.

Let $\bs \in \cS^{\infty}$ be a fixed state vector.
We divide $\bs$ into consecutive {\em chunks} according to the chunks of $\bp$ such that $\bs=\bs_1,\bs_2,\dots, \bs_\ell, \bs_{\ell+1}$ where for $i \in [\ell]$ the length of $\bs_i$ is $n_i$.
Here $n_1,\dots,n_{\ell-1}$ are defined by $\bp$, and $n_\ell$ will be defined shortly.
Let $\bp=\bp_1,\bp_2,\dots,\bp_{\ell},\bp_{\ell+1}$ be the corresponding chunk decomposition of $\bp$ in which, for $i \in [\ell]$, $\bp_i=p_i^{n_i}$, and $\bp_{\ell+1}=p_{\ell}^\infty$.
Let
{\small{$$
\tau({\bf s})=(1+\delta)\tau^{({\bf p})}_k({\bf s})=(1+\delta)\arg\min_\tau{\{\sum_{j<\tau}I_{{s_j}}(
X_{j},Y_j)\geq k\}},
$$}}
where $X_j$ is distributed according to the $j$'th entry of $\bp$, and $Y_j$ is the outcome of $W_{s_j}(\cdot|X_j)$ where $s_j$ is the $j$'th entry of $\bs$. 
Assume that $\tau(\bs) > \sum_{i=1}^{\ell-1}n_i$ and set $n_{\ell}=\tau(\bs)-\sum_{i=1}^{\ell-1}n_i$.
Namely, $\tau(\bs)=\sum_{i=1}^\ell n_i$.
We remove this assumption later at the end of the proof.
We employ the method of types in our decoding according to the definitions in \cite{CK97}.
For $i \in [\ell]$, let $q_{\bs_i} \in \cP(\cS)$ be the type of $\bs_i$.
In what follows we assume a function $f$ such that, for $i \in [\ell]$, we have $n_i \geq f(\delta)\tau(\bs)$ and $q_{\bs_i}(s) \geq f(\delta)$ for any $s \in \cS$.
This assumption is also removed at the end of the proof.

To simplify our presentation, for $i \in [\ell]$ and $s \in \cS$, we further partition each chunk $\bs_i$ to sub-chunks $\bs_{i,s}$ corresponding to entries in which $\bs_i$ equals $s$.
We assume without loss of generality that, for each $s \in \cS$, the sub-chunk $\bs_{i,s}$ consists of consecutive entries of $\bs_i$.
This follows from the fact that the decoder, knowing $\bs_i$, can  reorder the received information accordingly and from the fact that the random encoding rule is consistent over chunks.
Thus, we have, for $i \in [\ell]$ and $\cS=\{s_1,\dots,s_{|\mathcal S|}\}$, that $\bs_i=\bs_{i,s_1},\bs_{i,s_2},\dots,\bs_{i,s_{|\mathcal S|}}$.

Let $\bX=X_1,X_2,\dots$ be the vector of random variables corresponding to ${\bf p}$ in which, for any positive integer $j$, $X_j$ is distributed according to the $j$'th entry in $\bp$.
We divide $\bX$ into chunks as well corresponding to the decomposition of $\bp$ and $\bs$.
Namely, let $\bX=\bX_1,\bX_2,\dots,\bX_{\ell},\bX_{\ell+1}$ where, for $i \in [\ell]$, the length of $\bX_i$ is $n_i$. 
Let $\cC$ be a random code as discussed above.
Namely, for each message $m \in [2^k]$, the codeword corresponding to $m$ is independently distributed according to $\bX$ and will be denoted by $\bx(m)=\bx_1(m),\bx_2(m),\dots,\bx_{\ell}(m),\bx_{\ell+1}(m)$, where, for $i \in [\ell]$, the length of $\bx_i(m)$ is $n_i$ and each entry in $\bx_i(m)$ is distributed according to $p_i$.
Moreover, for $i \in [\ell]$, we divide $\bx_i(m)$ into sub-chunks according to those of $\bs_i$.
Namely, $\bx_i(m)=\bx_{i,s_1}(m),\bx_{i,s_2}(m),\dots,\bx_{i,s_{|\mathcal S|}}(m)$ where, for $s \in S$, the entries of $\bx_{i,s}(m)$ correspond to those in $\bs_{i,s}$.
Finally, let $\by=\by_1,\by_2,\dots,\by_{\ell},\by_{\ell+1}$ be the channel output at the receiver, and, for $i \in [\ell]$, let $\by_i=\by_{i,s_1},\by_{i,s_2},\dots,\by_{i,s_{|\mathcal S|}}$ be the chunk and sub-chunk decomposition respectively, both according to the chunk and sub-chunks of $\bx(m)$ and $\bs$.
It now holds, for $i \in [\ell]$, $s \in \cS$, and $m \in [2^k]$ that
each entry of $\bx_{i,s}$ is independently distributed according to $p_i$ and that 
$\by_{i,s}$ is distributed according to $W_s^{n_{i,s}}(\cdot|\bx_{i,s})$.
Here, $n_{i,s}$ is the length of the sub-chunk $\by_{i,s}$.
Note that, by our assumption on the type of $\bs_i$ and on the chunk size $n_i$, it holds that $n_{i,s} \geq f^2(\delta)\tau(\bs)$.
Namely, for the sub-chunk corresponding to $(i,s)$ our analysis reduces to the performance of the fixed memoryless channel $W_s$ of blocklength 
$n_{i,s}$ in the presence of a code generated randomly according to the fixed distribution $p_i$.
Let $I_{\min} = \min_{i,s}I_s(X,Y)$ and $I_{\max} = \max_{i,s}I_s(X,Y)$ for $(X,Y)$ distributed according to $p_i(x)W_s(y|x)$. 
Recall from the theorem statement that $I_{\min}>0$.
Notice, by our definition of $\tau^{({\bf p})}_k({\bf s})$, $I_{\min}$ and $I_{\max}$, that
$\frac{k}{I_{\min}} \geq \tau^{({\bf p})}_k({\bf s}) \geq \frac{k}{I_{\max}}$. 
We are now ready to define our encoder and decoder.

\textbf{Encoding:}
Let $\cC$ with encoder $E:[2^k]\rightarrow \cX^\infty$ be the randomly constructed code as defined above.
The encoder picks a uniform $m \in [2^k]$ and transmits $\bx(m)$.
We show below, for sufficiently large $k$, that with high probability the resulting code has marginals that are $\delta$-close to ${\bf p}$.
Let $p^i$ be the $i$'th entry of ${\bf p}$ (we use a superscript here to distinguish with $p_i$ which governs the $i$'th chunk of ${\bf p}$), and let $q^i$ be the marginal distribution of the $i$'th entry of the code $E$.
Using Sanov's theorem and Pinsker's inequality, for any entry $i$ in the codebook it holds with probability at most $2^{-\Omega{(2^k}\delta^2)}$ that ${\tt TV}(p^i,q^i) > \delta/2$. 
Thus, using standard concentration bounds and the independence between entries of $\cC$, it is not hard to verify that 
$\sum_{n=1}^\infty\Pr[\sum_{i=1}^n{\tt TV}(p^i,q^i)> \delta n] \leq 2^{-\Omega(k)}$.
In both bounds above, we use $k$ sufficiently large.
This implies that, for sufficiently large $k$, with probability at least $1-2^{-\Omega(k)}$ over code-design, the resulting code is has corresponding marginal distributions that are $\delta$-close to ${\bf p}$.

\textbf{Decoding:} 
Let $\by$ be the received transmission and $\bs$ the state sequence (that is available to the decoder).
Decoding follows standard typicality decoding outlined in the context of AVC's with DSI in, e.g., \cite{Stam:75} and Exercise 12.16(b) of \cite{CK97}.
Let $g(\delta)$ be a function of $\delta$ to be defined later.
Let $\tau(\bs)=\sum_{i=1}^{\ell}n_i=\sum_{i,s}n_{i,s}$.
The decoder iterates over $\hat{m}$ between $1$ and $2^k$, and decodes $\by^\tau$ to the first $\hat{m}$ such that for all $i \in [\ell]$ and $s \in \cS$ the joint type $q_{\bx_{i,s}(\hat{m}),\by_{i,s}} \in \cP(\cX) \times \cP(\cY)$ of the pair  $(\bx_{i,s}(\hat{m}),\by_{i,s})$ is of $\ell_\infty$ distance at most $g(\delta)$ from the distribution $q_{i,s}(x,y)=p_i(x)W_s(y|x)$ over $\cX \times \cY$.
If no message $m$ passes the test above, the decoder decodes arbitrarily to $m=1$.


We now prove the following concentration on $\cC$ for decoding with average error $\delta$.

\begin{claim}
    \label{claim:double}
    Let $\delta >0$.
    Let $i \in [\ell]$ and $s \in \cS$.
    With probability $2^{-\Omega(2^{k}/k)}$ over code design, it holds that 
    \begin{align*}
    \Pr_{m}&[\exists (i,s) \in [\ell] \times \cS \ \text{s.t.}\ 
     \|q_{\bx_{i,s}(m),\by_{i,s}}-q_{i,s}\|_\infty > g(\delta),
    \\
    &  \ \ \ \ \text{or}\ \ \exists m' < m, \forall (i,s), \|q_{\bx_{i,s}(m'),\by_{i,s}}-q_{i,s}\|_\infty \leq g(\delta)] \leq \frac{1}{k}
    \end{align*}
\end{claim}

\begin{IEEEproof}
For $m \in [2^k]$, let $Z(m)$ be the indicator of the error event that either 
$\exists (i,s)$ such that $\|q_{\bx_{i,s}(m),\by_{i,s}}-q_{i,s}\|_\infty > g(\delta)$ or 
$\exists m' < m$ such that $\forall (i,s), \|q_{\bx_{i,s}(m'),\by_{i,s}}-q_{i,s}\|_\infty \leq g(\delta)$.
Let $\cC^{(m-1)}$ be any fixed values for the codewords $\bx(1),\dots,\bx(m-1)$.
We first analyze the expected value of $Z(m)$ conditioned on $\cC^{(m-1)}$.

Standard analysis, appearing for example in Chapter 7 of \cite{yeung2008information}, implies the existence of $\hat{g}(\delta)$ that tends to zero when $g(\delta)$ tends to zero, such  that for $m$ and any $(i,s)$,
$$
\Pr_{\mathcal C}[\|q_{\bx_{i,s}(m),\by_{i,s}}-q_{i,s}\|_\infty > g(\delta) \mid \cC^{(m-1)}]
\leq 2^{-n_{i,s}\hat{g}(\delta)}.
$$
and for any $m'<m$ and any $(i,s)$,
$$
\Pr_\cC[\|q_{\bx_{i,s}(m'),\by_{i,s}}-q_{i,s}\|_\infty \leq g(\delta) \mid \cC^{(m-1)}]
\leq 2^{-n_{i,s}(I(X_{i};Y_s)-\hat{g}(\delta))}.
$$
Here $(X_i,Y_s)$ are distributed according to $p_i(x)W_s(y|x)$.
As the events in every sub-chunk are independent, we have
for any $m'<m$ that,
\begin{align*}
\Pr_\cC[\forall (i,s), \|q_{\bx_{i,s}(m'),\by_{i,s}}-&q_{i,s}\|_\infty \leq g(\delta) \mid \cC^{(m-1)}]\\
& \leq 
2^{-\sum_{i,s}n_{i,s}(I(X_{i};Y_s)-\hat{g}(\delta))}.
\end{align*}
Thus, by the union bound (over $m'<m$ and more), 
\begin{align*}
    \Pr_\cC[&Z(m)=1 \mid \cC^{(m-1)}] \\
    &\leq 
    2^k2^{-\sum_{i,s}n_{i,s}(I(X_{i};Y_s)-\hat{g}(\delta))}  + \sum_{i,s}2^{-\hat{g}(\delta)\min_{i,s}n_{i,s}}.
\end{align*}
As, $\tau(\bs)=\sum_{i,s}n_{i,s}=(1+\delta)\tau_{{k}}^{(\bp)}(\bs)$ it holds that
$\sum_{i,s}n_{i,s}I(X_i;Y_s)\geq k+\delta \tau_{{k}}^{(\bp)}(\bs)I_{\min} \geq k\left(1+\frac{\delta I_{\min}}{I_{\max}}\right)$.
We now have that 
\begin{align*}
    E_{\cC}[&Z(m) \mid \cC^{(m-1)}] = \Pr_{\cC}[Z(m)=1 \mid \cC^{(m-1)}] \\
    & \leq 
    2^k2^{-k-k\frac{\delta I_{\min}}{I_{\max}}+\hat{g}(\delta)\tau(\bs)}
    + \sum_{i,s}2^{-\hat{g}(\delta)\min_{i,s}n_{i,s}}\\
    & \leq 
    2^{-k\left(\frac{\delta I_{\min}}{I_{\max}}-\frac{\hat{g}(\delta)}{I_{\min}}\right)}+
    \ell |\mathcal S|2^{-k\frac{\hat{g}(\delta)f^2(
\delta)}{I_{\max}}}
\leq \frac{1}{2k},
\end{align*}
for $\delta$ and thus $g(\delta)$ sufficiently small such that $\hat{g}(\delta)\leq\frac{\delta I_{\min}}{2I_{\max}}$ and $k$ sufficiently large. 

Let $Z=\sum_{m=1}^{2^k}Z(m)$.
To conclude the claim assertion, using the Chernoff-type Lemma A.1 of \cite{CsiszarN:88positivity} and noting that $Z(m)$ depends only on $\bx_1,\dots,\bx_m$, we have
$$
\Pr_\cC\left[Z \geq \frac{2^{k}}{k} \right] \leq 2^{-\Omega(2^{k}/k)}.
$$
\end{IEEEproof}

Consider any $\bs \in \cS^\infty$.
Let $\tau_{\max} = \max_{\bs}\tau(\bs) = \frac{(1-\delta)k}{I_{\min}}$.
Thus, using a union bound over all $\bs \in \cS^\infty$ of length at most $\tau_{\max}$, Claim~\ref{claim:double} implies the existence of a code $\cC$ that is $\delta$-close to ${\bf p}$ and a DSI-decoder that for any $\bs$ decodes at time $\tau(\bs)$ with average error at most $1/k$. 
We conclude that the competitive ratio obtained by the suggested scheme is 
\begin{align}
\label{eq:ach_end}
    \inf_{\bs}\frac{\tau^*_k({\bs})}{\tau(\bs)}
    = \inf_{\bs}\frac{\tau^*_k{(\bs)}}{(1+\delta)\tau_{{k}}^{(\bp)}(\bs)}
    \geq (1-\delta)\inf_{\bs}
    \left(\frac{\tau^*_k({\bf s})}{\tau^{({\bf p})}_k({\bf s})}\right).
\end{align}

To conclude our proof, we revisit the functions $f$, $g$ and $\hat{g}$ alongside our assumptions.
Throughout, we fix $\delta$ to be a sufficiently small constant and $k$ asymptotically large. 
We chose $f(\delta) \rightarrow 0$ as $\delta \rightarrow 0$.
We chose $g(\delta) \rightarrow 0$ as $\delta \rightarrow 0$ to be a sufficiently small function of $\delta$ to satisfy the requirement $\hat{g}(\delta)\leq\frac{\delta I_{\min}}{2I_{\max}}$ stated above.
If we remove the assumptions that for all $i \in [\ell]$, $n_i \geq f(\delta)\tau(\bs)$ and for all $s \in \cS$, $q_{{\bf s}_i}(s)\geq f(\delta)$ then the decoder can neglect the entries corresponding to {\em bad} pairs $(i,s)$ that violate (one of) the assumptions.
The cumulative mutual information $\sum_{i,s}n_{i,s}I
(X_i,Y_s)$ (summed over bad pairs) lost at the decoder is bounded by $\tau(\bs)I_{\max}(\ell f(\delta)+ |\mathcal S|f(\delta))$ which for sufficiently small $\delta$ and suitable $f(\delta)$ leaves the remaining cumulative mutual information $\sum_{i,s}n_{i,s}I
(X_i,Y_s)$ (summed over good pairs) at the decoder to be at least  $k\left(1+\frac{\delta I_{\min}}{2I_{\max}}\right)$
which replaces $\sum_{i,s}n_{i,s}I
(X_i,Y_s) \geq k\left(1+\frac{\delta I_{\min}}{I_{\max}}\right)$ in the previous presented proof.
If the assumption that $\tau(\bs) > \sum_{i=1}^{\ell-1}n_i$ does not hold, and instead it holds that $\sum_{i=1}^{\ell^*-1}n_i < \tau(\bs) \leq \sum_{i=1}^{\ell^*}n_i$ then one need only consider the prefix of $\bp$ consisting of the first $\ell^*$ chunks and replace $\ell$ in the analysis by $\ell^*$. 


 \section{Proof of $\CR\le  1/2$}
In this section, we prove that for the family of channels in Fig.~\ref{fig:avc_example} we have $\CR\le 1/2$. The main idea is to identify from the code analysis of our lower bound a collection $\widehat{\mathcal S}^\infty \subseteq \mathcal S^\infty$ of state sequences that constrain the optimization defining the competitive ratio $\CR$. 
Namely, 
\begin{align}
\label{eq:AppC}
        {\CR} &\le 
        \lim_{k\to\infty}\max_{\bp \in \mathcal P^\infty} \min_{\bs\in \mathcal S^{\infty}} \left(\frac{\tau^*_k({\bf s})}{\tau^{({\bf p})}_k({\bf s})}\right)\nn\\
        &\le 
    \lim_{k\to\infty}\max_{\bp \in \mathcal P^\infty} \min_{\bs\in  \widehat{\mathcal S}^{\infty}} \left(\frac{\tau^*_k({\bf s})}{\tau^{({\bf p})}_k({\bf s})}\right). 
\end{align}
Bounding $\CR$ by the top expression in \eqref{eq:AppC} follows from standard capacity bounds, e.g., \cite[Section VI]{LangbergSabag24}.
The upper bound in the bottom inequality of \eqref{eq:AppC} holds for any subset $\widehat{\mathcal S}^\infty$  of state sequences.
Here, we consider $\widehat{\mathcal S}^\infty$ consisting of two sets.
The first is $\widehat{\mathcal S}_1 = \{1^ks^\infty\}$ corresponding to the two sequences whose prefix of length $k$ is $1^k$, followed by a constant-state sequence $s^\infty$ with $s\in\{1,2\}$. 
The second set is defined as $\widehat{\mathcal S}_2 = \{2^k\bs_{1:2}^{3k/4}s^\infty\}$ where $\bs_{1:2}^{3k/4}$ is any sequence of length $3k/4$ whose number of states of type $1$ and $2$ satisfy a ratio of $[1:2]$, e.g., $1^{k/4}2^{k/2}\in \bs_{1:2}^{3k/4}$. Simply put, we fix the type of the sequence in this interval. 

The particular choice of these sets is based on their optimal decoding times. We have $\tau^*_k(\bs)=k$ for all $\bs\in\widehat{\mathcal S}_1$ since their prefix is $1^k$. For the second set, we have $\tau^*_k(\bs)=7k/4$ for all $\bs\in\widehat{\mathcal S}_2$ since the location of $1$'s in the interval $[k,7k/4]$ has no impact on the optimal decoding time.

The remainder of the proof consists of two main steps. The first step is to show that the optimization of input distributions in \eqref{eq:AppC} can be limited to input distributions that are constant within the intervals $(0,k], (k,7k/4], (7k/4,\infty)$. This step will follow from the structure of the chosen state-sequence sets $\widehat{\mathcal S}_1$ and $\widehat{\mathcal S}_2$. Then, to complete the proof of the upper bound, in the second step we  compute the optimization in \eqref{eq:AppC} under the restricted domain of input distributions.

Our first claim is that during the time interval $[1,k]$ the optimizing input distribution is fixed and need not change.
This follows from the fact that no matter which state sequence is realized, the optimal stopping time is at least $k$.
Thus, by the concavity of mutual information, any collection of time-varying input distributions in this interval is sub-optimal.
The fixed distribution in this interval is denoted by $p_1\in\cP(\cX)$. 

For the second interval, $[k,7k/4]$, we also claim that a fixed input distribution optimizes the competitive ratio.
The argument follows from symmetry: the type of the state sequence is fixed, thus we can take an expected value over the state sequence to upper bound the minimal cumulative mutual information, i.e., 
\begin{align}
    & \min_{\bs\in\bs_{1:2}^{3k/4}} \sum_{i=1}^{3k/4} I_{s_i}(X_i;Y_i) = \min_{\bs\in\bs_{1:2}^{3k/4}} \sum_{i=1}^{3k/4} \sum_{s =1,2} \mathbb{1}\{s_i=s\} I_{s}(X_i;Y_i)\nn\\
    &\stackrel{(a)}\le \mathbb{E}_{S^{3k/4}}\left[ \sum_{i=1}^{3k/4} \sum_{s=1,2} \mathbb{1}\{S_i=s\}I_{s}(X_i;Y_i)\right] \nn\\
    &\stackrel{(b)}= \frac{3k}{4} \mathbb{E}_{S^{3k/4},T}\left[ \sum_{s=1,2} \mathbb{1}\{S_T=s\}I_{s}(X_T;Y_T|T)\right], 
\end{align}
where in step $(a)$ we bound the minimum sequence by a uniform distribution over a uniform random sequence $S^{3k/4}$ of the corresponding fixed type. In step $(b)$, we define a uniform random variable $T\sim\text{U}(1:3k/4)$, independent of $S^{3k/4}$. 
The latter expectation converges for large $k$ to $I(X;Y|S,T)$ with $S\sim \text{Bern}(1/3)$ and $(T,S,X,Y)$ distributed according to  $p(t)p(s)p(x|t)p(y|x,s)$. We note that $ I(X;Y|S,T)\le I(X;Y|S)$ where the joint distribution on the right hand side is $p(s)p(x)p(y|x,s)$ with $p(x) = \sum_t p(t)p(x|t)$.
The inequality follows from the Markov chain $T-(X,S)-Y$. 
To conclude, an i.i.d. distribution of channel inputs will achieve the upper bound for the cumulative mutual information for large $k$. 
A similar result holds in the interval $[k,7k/4]$ for 
$\widehat{\mathcal S}_1$ 
as well since the state is constant during this interval. In particular, this follows from the concavity of mutual information.

Finally, we note that since for any $\bs$, $\tau^*_k(\bs)\leq 7k/4$, and since our sets of sequences are symmetric 
after time $7k/4$, 
the best rate after time $7k/4$ is achieved when the rates of both channels are equal, i.e., the best rate is $\frac{1}{3}$. This completes the first step of the proof, to show that the optimal input distributions can be restricted to be constant in the specified intervals.



We proceed with the second step of our proof in which we compute the stopping times induced by each set of sequences. From the stopping times we obtain values for competitive ratios of the form $\frac{\tau^*_k({\bf s})}{\tau^{({\bf p})}_k({\bf s})}$; then, optimizing over these ratios according to \eqref{eq:AppC} will result in our upper bound on the competitive ratio.

For sequences in $\bs\in\widehat{\mathcal S}_2$, we have
\begin{align}
    k&= k \frac{p_1}{2} + \frac{k}{4}(1-p_2) + \frac{k}{2}\frac{p_2}{2} + \frac1{3}\left(\tau_k^{({\bf p})}(\bs)-\frac{7}{4}\right)\\
    &= k\frac{p_1}{2} + \frac{k}{4} + \frac1{3}\left(\tau_k^{({\bf p})}(\bs) - \frac{7k}{4}\right).
\end{align}
Note that the stopping time is independent of $p_2$. This is due to our choice of the ratio $[1:2]$ in the definition of $\widehat{\mathcal S}_2$. We obtain that the stopping time is $\tau_k^{({\bf p})}(\bs) = 4k - \frac{3kp_1}{2}$ and that the competitive ratio for this set is $\frac{\tau_k^\ast(\bs)}{\tau_k^{({\bf p})}(\bs)} = \frac{7/4}{4k - \frac{3kp_1}{2}}$.

For the set $\bs\in\widehat{\mathcal S}_1$, we know that the stopping time is greater than $k$ but we need to consider different cases depending on whether the stopping time is smaller or greater than $7k/4$. Recall that there are two sequences in $\widehat{\mathcal S}_1 = 1^ks^\infty$ and therefore we have four cases. We start by computing the stopping time of any sequence in $\widehat{\mathcal S}_1$ such that $\tau_k^{({\bf p})}(\bs) \ge \frac{7k}{4}$:
\begin{align}
k &= (1-p_1)k + \frac{3k}{4} R(s,p_2) + \frac1{3}\left(\tau_k^{({\bf p})}(\bs) - \frac{7k}{4}\right),
\end{align}
where $R(1,p_2) = 1-p_2$ and $R(2,p_2) = p_2/2$ depending on whether the sequence in $\widehat{\mathcal S}_1$ ends with $\bs^\infty=1^\infty$ or $\bs^\infty=2^\infty$. Simplifying the equation, we obtain that the stopping time is $\tau_k^{({\bf p})}(\bs) = \frac{7k}{4} + 3kp_1 - \frac{9k}{4}R(s,p_2)$ subject to input distributions that satisfy 
$3p_1 - \frac{9}{4}R(s,p_2)\ge0$ (or equivalently, $\frac{p_1}{R(s,p_2)}\ge \frac{3}{4}$). Moreover, the competitive ratio is $\frac1{\frac{7}{4} + 3p_1 - \frac{9}{4}R(s,p_2)}$.

We next consider the stopping time of sequences in $\widehat{\mathcal S}_1$ with $\tau_k^{({\bf p})}(\bs) \le \frac{7k}{4}$:
\begin{align}
k &= (1-p_1)k +  R(s,p_2)\left(\tau_k^{({\bf p})}(\bs) - k\right),
\end{align}
which provides with the stopping time $\tau_k^{({\bf p})}(\bs)= k + \frac{p_1k}{R(s,p_2)}$, subject to input distributions that satisfy $\frac{p_1}{R(s,p_2)}\le \frac{3}{4}$.

To determine the competitive ratio for $\widehat{\mathcal S}_1$, we analyze four different cases. 
\begin{enumerate}
    \item Case A: $\tau_k(\bs)\ge 7/4$ for both sequences in $\widehat{\mathcal S}_1$. 
    \item Case B: For $\bs^\infty=1^\infty$, we have $\tau_k(\bs)\ge 7/4$, while for $\bs^\infty=2^\infty$ we have $\tau_k(\bs)\le 7/4$. 
    \item Case C: $\tau_k(\bs)\le 7/4$ for both sequences in $\widehat{\mathcal S}_1$.
    \item Case D: For $\bs^\infty=1^\infty$, we have $\tau_k(\bs)\le 7/4$, while for $\bs^\infty=2^\infty$ we have $\tau_k(\bs)\ge 7/4$. 
\end{enumerate}
One can derive the conditions on $p_1,p_2$ that specify each of the above cases. For example, Case A is valid only if $\{3p_1 - \frac{9}{4} (1-p_2)\ge0\}\cap\{3p_1-\frac{9}{4}\frac{p_2}{2}\ge0 \}$. We omit the derivation of the other cases and proceed to directly compute an upper bound on the competitive ratio conditioned on each case. The upper bound on the overall competitive ratio is the largest upper bound among the different cases and will be shown to be upper bounded by $0.5$.

For Case A, we have that the optimized competitive ratio is 
\begin{align}\label{eq:proof_ub_A}
    &\max_{p_1\ge1/4} \max_{1-\frac{4p_1}{3}\le p_2\le \frac{8p_1}{3}} \nonumber\\
    &\min\left\{\frac1{\frac{7}{4} + 3p_1 - \frac{9}{4}\min_s\{R(s,p_2)\}}, \frac{\frac{7}{4}}{4-3p_1/2}\right\}\nonumber\\
    &= \max_{p_1}\min\left\{\frac1{1 + 3p_1}, \frac{\frac{7}{4}}{4-3p_1/2}\right\}\nn\\
    &= 1/2,
\end{align}
where the first equality follows by the fact that the second term does not depend on $p_2$; thus, optimizing over the first expression yields $p^*_2 = 2/3$. The second equality is obtained by comparing the two terms (which are equal when $p_1^*  = 1/3$). Indeed, the optimal point here lies in the feasible region so we have equalities in both steps. 
We now turn to Cases B-D, and show, as well, that the induced upper bound is less or equal than $1/2$. 

For Case B, we have $\tau_k(\bs)\ge\frac{7}{4}$ for $s=1$ but $\tau_k(\bs)\le\frac{7}{4}$ for $s=2$. Combining these competitive ratios with $\widehat{S}_2$ yields 
\begin{align}
    &\max_{p_1\le \frac{3}{8}} \max_{\max\{1-\frac{4p_1}{3},\frac{8p_1}{3}\}\le p_2} \nonumber\\
    &\min\left\{\frac1{\frac{7}{4} + 3p_1 - \frac{9}{4}(1-p_2)}, \frac{1}{1 + \frac{2p_1}{p_2}},\frac{\frac{7}{4}}{4-3p_1/2}\right\}\nonumber\\
    &\stackrel{(a)}\le \max_{p_1\le \frac{3}{8}} \min\left\{ \frac{1}{1 + 3p_1},\frac{\frac{7}{4}}{4-3p_1/2}\right\}\nonumber\\
    &\stackrel{(b)}\le 0.5,
\end{align}
where $(a)$ follows by comparing the two terms that depend on $p_2$. 
These two terms are equal when $p_2 = 2/3$ or $-4p_1/3$, and it is clear that $p_2 = -4p_1/3$ is not feasible so we choose $p_2=2/3$. Note that $(a)$ is an upper bound (and not equality) since $p_2=2/3$ may not be feasible depending on the value of $p_1$. Step $(b)$ follows from \eqref{eq:proof_ub_A} by ignoring the constraint on $p_1$.  

For Case C, we have 
\begin{align}
    &\max_{p_1\le \frac{1}{4}} \max_{\frac{8p_1}{3}\le p_2\le 1-\frac{4p_1}{3}} \min\left\{\frac{1}{1 + \frac{p_1}{R(s,p_2)}},\frac{\frac{7}{4}}{4-3p_1/2}\right\}\nonumber\\
    &\le 0.5.
\end{align}
The upper bound is simple since $p_2 = 2/3$ that equates both terms depending on $p_2$ lies in the feasible region. We then maximize over $p_1\in[0,1]$ that gives $1/2$ as in Case A.

For last case, Case D, the optimized competitive ratio is
\begin{align}
    &\max_{p_1\ge \frac{3}{4}} \max_{p_2 \le \min\{1-\frac{4p_1}{3},\frac{8p_1}{3}\}} \nonumber\\
    &\min\left\{\frac1{\frac{7}{4} + 3p_1 - \frac{9}{8}p_2},\frac{1}{1 + \frac{p_1}{1-p_2}},\frac{\frac{7}{4}}{4-3p_1/2}\right\},
\end{align}
and comparing the terms depending on $p_2$ provides $p^*_2\in \{2/3,1+8p_1/3\}$. The only feasible point is $p_2=2/3$ which simplifies the optimization to be as in Case A. We note that the optimal unconstrained solution does not lie in the feasible region and therefore $0.5$ is a strict upper bound.

Combining the different cases, we conclude that the competitive ratio is upper bounded by the maximum among these bounds and is thus equal to $0.5$ as asserted.
\end{document}